\documentclass[12pt]{article}
\font\mybb=msbm10 at 12pt
\def\bb#1{\hbox{\mybb#1}}
\begin{document}
\textheight=24cm
\textwidth=16.5cm
\topmargin=-1.5cm
\oddsidemargin=-0.25cm.

\begin{flushright}
{Contribution to SQS 2017 Proceedings.  
ISSN 1063-7796, \\ Phys. Part. Nucl., 2018, Vol. 49, No. 5, pp. 829-834. 
}
\end{flushright}

\vspace{3cm}

\begin{center}
{\Large\bf On 10D SYM superamplitudes}

\bigskip
{Igor Bandos$^{\dagger\ddagger}$}
\bigskip \\
{\it $^{\dagger}$ Department of
Theoretical Physics, University of the Basque Country UPV/EHU,
P.O. Box 644, 48080 Bilbao, Spain.
 \\
$^{\ddagger}$ IKERBASQUE, Basque Foundation for Science, 48011, Bilbao, Spain
 }
\end{center}

\bigskip

{\bf Abstract:} {\rm Recently the spinor helicity and (two types of) superamplitude formalisms for 11D supergravity and 10D supersymmetric Yang-Mills theories were proposed in
\cite{Bandos:2016tsm,Bandos:2017eof,Bandos:2017zap}. In this contribution we describe  briefly the basic properties of these superamplitudes for the simpler case of 10D SYM.
}

\bigskip

{\bf 1.} {\it Introduction}. The impressive  recent progress in calculation of loop  amplitudes in maximally  supersymmetric $D=4$ gauge theory and supergravity, ${\cal N}=4$ SYM and
${\cal N}=8$ SUGRA, has been reached in the frame of on--shell amplitude calculus (see \cite{Bern:2011qn,Benincasa:2013faa,Elvang:2015rqa,ArkaniHamed:2017} and refs. therein)
using intensively the so-called spinor helicity formalism and on-shell superfield approach.

These are based on the description of massless particle momentum and polarizations by single complex Weyl
spinor $\lambda_{\alpha}$ ($\alpha=1,2$) called helicity spinor, and its complex conjugate $\bar{\lambda}_{\dot{\alpha}}$ ($\dot{\alpha}=1,2$). The (light-like) momentum $p_{\mu (i)}$ ($\mu=0,1,2,3$) of $i$-th particle is given by
\begin{eqnarray}\label{p=ll=4D}
p_{\alpha\dot{\alpha} (i)}:= p_{\mu (i)} \sigma^\mu_{\alpha\dot{\alpha}}=  2\lambda_{\alpha(i)}\bar{\lambda}{}_{\dot{\alpha}(i)}   \quad \Leftrightarrow \quad p_{\mu(i)}= \lambda_{(i)} \sigma_\mu \bar{\lambda}_{(i)},  \qquad
\end{eqnarray}
where $\sigma^\mu_{\alpha\dot{\alpha}}$ are relativistic Pauli matrices. The polarization vectors of spin 1  particles ('gluons') of negative and positive helicity are
\begin{eqnarray}\label{polar=}
\varepsilon^{(+)}_{\alpha\dot{\alpha}(i)}= {\lambda_{\alpha(i)}\bar{\mu}{}_{\dot{\alpha}} }/{[\bar{\lambda}_{(i)}\bar{\mu}]}  \; ,  \qquad \varepsilon^{(-)}_{\alpha\dot{\alpha}(i)}=  {\mu_{\alpha}\bar{\lambda}{}_{\dot{\alpha}(i)}}/ {<\mu\lambda_{(i)}>}
\; ,   \qquad  \end{eqnarray}
where $\bar{\mu}{}_{{\dot{\alpha}}}= (\mu_{\alpha})^*$ is a (constant) reference spinor,
and 
\begin{eqnarray}\label{lamu=}
& <\mu\lambda_{(i)}>:= \epsilon_{\alpha\beta} \mu^\alpha\lambda^\beta_{(i)} \equiv  \mu^2\lambda^1_{(i)} -  \mu^1\lambda^2_{(i)}\, , \qquad {}[\bar{\mu}\bar{\lambda}_{(i)}]=
\epsilon_{{\dot{\alpha}\dot{\beta}}} \bar{\mu}^{{\dot{\alpha}}}\bar{\lambda}^{{\dot{\beta}}}_{(i)}\; .   \qquad  \end{eqnarray}

The $n$-point scattering amplitudes, which we define with explicitly extracted momentum conservation delta function,
\begin{eqnarray}\label{cAn=cAnl}
\delta^4 (\sum\limits_{i=1}^{n} p_{i}) {\cal A}(p_{1},\varepsilon_{1}; ...;p_{n},\varepsilon_{n}) =
{\cal A}(\lambda_{1}, \bar{\lambda}_{1}; \ldots ;\lambda_{n}, \bar{\lambda}_{n}) \delta^4 (\sum\limits_{i=1}^{n} \lambda_{\alpha i} \bar{\lambda}_{\dot{\alpha}i})\; , \;
 \end{eqnarray}
 are independent of the choice of $\mu$ in (\ref{polar=}), and obey the helicity constraints,
\begin{eqnarray}\label{hhiA=hiA}
\hat{h}_{(i)} {\cal A}(1,...,n) = h_i {\cal A}(1,...,n)\; .\qquad
\end{eqnarray}
Here $h_i$ is the helicity of the state, $h_i=\pm 1$ in the case of gluons, and
\begin{eqnarray}
 \label{Ui=b}
& \hat{h}_{(i)}:=\frac 1 2 \lambda^\alpha_{(i)} \frac \partial  {\partial \lambda^\alpha_{(i)} } -
\frac 1 2 \bar{\lambda}{}^{{\dot{\alpha}}}_{(i)} \frac \partial  {\partial \bar{\lambda}{}^{{\dot{\alpha}}}_{(i)}} \,   \qquad \end{eqnarray}
is the  helicity operator. Eq. (\ref{hhiA=hiA})  implies
 \begin{eqnarray}\label{cA=e2ihicA}
 {\cal A} (..., e^{i\beta_i} \lambda^\alpha_{(i)},  e^{-i\beta_i}  \bar{\lambda}{}^{\dot{\alpha}}_{(i)}, ...) = e^{2ih_i \beta_i } {\cal A}(..., \lambda^\alpha_{(i)},   \bar{\lambda}{}^{\dot{\alpha}}_{(i)}, ...)\; . \qquad
\end{eqnarray}

A {\it superamplitude}  of ${\cal N}=4$ SYM   depends, besides $n$ sets of complex bosonic spinors, on $n$ sets of complex fermionic variables $\eta_{Ai}$ 
  carrying the index of fundamental representation {\bf 4} of the $SU(4)$ R-symmetry group
  \begin{eqnarray}\label{cAn=cAnlf}
{\cal A}(1; ...;n) = {\cal A}(\lambda_{(1)}, \bar{\lambda}_{(1)}, \eta_{1}; ...;\lambda_{(n)}, \bar{\lambda}_{(n)},  \eta_{n}), \quad \eta_{iA}\eta_{Bj}=- \eta_{Bj}\eta_{Ai}
;    \; \end{eqnarray}
it obeys $n$ super-helicity constraints,
  \begin{eqnarray}\label{UcAn=2cAn}
 &  \hat{h}_{(i)} {\cal A}(\{\lambda_{(i)}, \bar{\lambda}_{(i)}, \eta_{Ai}\}) = {\cal A}(\{\lambda_{(i)}, \bar{\lambda}_{(i)}, \eta_{Ai}\})\; ,\qquad
\\ \label{Ui=}
&  \hat{h}_{(i)}=
 \frac 1 2 \lambda^\alpha_{(i)} \frac \partial  {\partial \lambda^\alpha_{(i)} }-\frac 1 2
 \bar{\lambda}{}^{\dot{\alpha}}_{(i)} \frac \partial  {\partial \bar{\lambda}{}^{\dot{\alpha}}_{(i)}}  +
 \frac 1 2  \eta_{Ai} \frac \partial  {\partial  \eta_{Ai} } \, ,  \qquad A=1,...,4\, .   \qquad
 \end{eqnarray}
The dependence of the superamplitude on $\eta_{Ai}$ is holomorphic:  it  is independent of $\bar{\eta}{}^A_{i}=(\eta_{Ai}) ^*$. Furthermore, according to (\ref{Ui=}),  the degrees of homogeneity in $\eta_{Ai}$ is related to the helicity $h_i$ in (\ref{Ui=b}),  so that the  decomposition of superamplitude on  $\eta_{Ai}$ includes amplitudes of different helicities.

These superamplitudes can be regarded as multiparticle generalizations of the so-called on-shell superfields
 \begin{eqnarray}\label{Phi=3,4}
 & \Phi (\lambda ,\bar{\lambda}, \eta^A) = f^{(+)} +  \eta_A \chi^A + \frac 1 2 \eta_B \eta_A s^{AB}+
  \eta^{\wedge 3\, A}{} \bar{\chi}_A + \eta^{\wedge 4}  f^{(-)} \; , \qquad
 \\
 \label{etaN:=}
& \eta^{\wedge 3\; A}= \frac 1 {3!} \epsilon^{AB_2B_3B_4} \eta_{B_2}\eta_{B_3}\eta_{B_{4}}
  \; , \qquad  \eta^{\wedge 4}=  \frac 1 {4!} \eta_{A_1} \ldots \eta_{A_{4}}
 \epsilon^{{A_1}  \ldots A_{4} }\; , \qquad
\end{eqnarray}
which  obey the  super-helicity constraint
 \begin{eqnarray}\label{UPhi=1}
 & \hat{h}  \Phi (\lambda ,\bar{\lambda}, \eta) =
 \Phi (\lambda ,\bar{\lambda}, \eta ) \; , \qquad 2\hat{h}=
 \lambda^\alpha \frac \partial  {\partial \lambda^\alpha }-
 \bar{\lambda}{}^{\dot{\alpha}}\frac \partial  {\partial \bar{\lambda}{}^{\dot{\alpha}}}  +
 \eta_A \frac {\partial }  {\partial  \eta_A} \,  .  \qquad
 \end{eqnarray}
The component fields in (\ref{Phi=3,4}) describe the on-shell degrees of freedom of the ${\cal N}=4$ SYM multiplet:
positive and negative helicity gluons ($f^{\pm}$), 6 scalars ($s^{AB}=-s^{BA}$) and four  $\pm 1/2$ helicity fermions ($\chi^A$ and $\tilde{\chi}{}_A$).

Higher $n>3$ (super)amplitudes can be reconstructed from the lower, $n'$ point (super)amplitudes with $3\leq n'\leq (n-1)$, using the BCFW recurrent relations \cite{Britto:2005fq} and its superfield generalization \cite{ArkaniHamed:2008gz}. To start such calculations one needs to know  the basic MHV and anti-MHV ($\overline{{\rm MHV}}$) 3-point superamplitudes which in the case of ${\cal N}=4$ SYM read
   \begin{eqnarray}
     \label{cA3g=MHV}
 {\cal A}^{MHV}(1,2, 3)=  \frac {1} {[12] \,[23]\, [31]}\; \delta^8 \left( \bar{\lambda}_{ \dot{\alpha}1}\eta_{A1}+  \bar{\lambda}_{ \dot{\alpha}2}\eta_{A2}+ \bar{\lambda}_{ \dot{\alpha}3} {\eta}_{A3} \right) ,  \qquad
\\    \label{cA3g=bMHV}
 {\cal A}^{\overline{{\rm MHV}}}(1,2, 3)= \frac { \delta^4 \left( \eta_{A(1)}<23> +  \eta_{A(2)}<31>+  \eta_{A(3)} <12> \right) }  {<12> <23><31>}\; .   \qquad
  \end{eqnarray}

 {\bf 2.}  {\it Analytic superamplitudes of 10D SYM and spinor moving frame nature of 10D spinor helicity variables.}
  {An approach to superamplitudes of 10D SYM}, which has a similarity with the above described 4D formalism have been developed in
 \cite{Bandos:2017zap}. In it the superamplitudes also depend analytically on a similar set of complex $4$ component variables
 $\eta^-_{A i}$, $A=1,...,4$, $i=1,...,n$. However, the amplitudes in these superamplitudes
   \begin{eqnarray}\label{cA=alt}
 {\cal A}_n\; \delta^D \left(\sum\limits_i^n k_{ai} \right) = {\cal A}_n (\{ \rho^\#_{i}, v_{\alpha qi}^{\; -}; w_i, \bar{w}_i; \eta_{A i}\}) \; \delta^D \left(\sum_i\limits^n \rho^{\#}_{i} u_{a i}^= \right)\;   \qquad \end{eqnarray}
depend on a different set of spinor helicity variables. This includes a set of 10D spinor frame (Lorentz harmonics) variables
$ v_{\alpha qi}^{\; -}$, which we are going to describe now, densities $ \rho^\#_{i}$, and a set of internal harmonic variables
$ w_i, \bar{w}_i$ (see \cite{Galperin:2001uw} and refs. therein) parametrizing the coset of $SO(8)$:
\begin{eqnarray}\label{wi=G-H=alt}
& \{ w^{A}_{qi}, \bar{w}{}_{Aqi}\} = \left(\frac {SO(8)}{SU(4)\otimes U(1)}\right)_i \; , \qquad q=1,...,8; \; , \qquad A=1,...,4\; . \qquad \end{eqnarray}
In (\ref{cA=alt}) these are actually pure gauge and play an auxiliary role  (see below).

{\bf 2.1.} {\it Spinor frame variables \cite{Bandos:1996ju,Bandos:2017eof,Bandos:2017zap} or Lorentz harmonics}, which are  suitable to describe $D=10$ massless particles \cite{Galperin:1991gk,Delduc:1991ir}, are given by the set of 8 strongly constrained  16-component real bosonic spinors $v_{\alpha q}^{\; -}$ defined up to $SO(1,1)\times SO(8)$ transformations. These constraints and identifications
allow to consider  $v_{\alpha q}^{\; -}$ as a kind of homogeneous coordinates of the eight-sphere
\begin{eqnarray}\label{v-=S8}
 \{ v_{\alpha q}^{\; -}\} = {\bb S}^{8}\; , \qquad \alpha=1,...,16\; , \qquad q=1,...,8\; .
\end{eqnarray}

The sets of such variables can be used to write the expressions for light-like momenta of a massless 10D particles similar to (\ref{p=ll=4D}),
\begin{eqnarray}\label{k=pv-v-11}
k_{ ai}
\sigma^a_{\alpha\beta}= 2\rho_{(i)}^{\#} v_{\alpha q(i)}^{\; -} v_{\beta q(i)}^{\; -}  \; , \qquad
 \rho^{\#} v^-_{\alpha {q}i} \tilde{\sigma}^{\alpha\beta}_{a}v^-_{\beta {p}i}= k_{a (i)} \delta_{{q}{p}}\; . \qquad
\end{eqnarray}
Here $\alpha,\beta=1,...,16$ are 10D Majorana-Weyl (MW) spinor indices and
$\sigma^a_{\alpha\beta}=\sigma^a_{\beta\alpha}$ and $\tilde{\sigma}^{a \; \alpha\beta}=\tilde{\sigma}^{a \;\beta\alpha}$ are 16$\times 16$ generalized Pauli matrices,
\begin{eqnarray}\label{sts+=}\sigma^a\tilde{\sigma}^b +\sigma^b\tilde{\sigma}^a=
2\eta^{ab}{\bb I}_{16\times 16}\; ,  \qquad a,b=0,1,...,9\; .
\end{eqnarray}

The constraints on $ v_{\alpha q(i)}^{\; -}$ are essentially given by the relations (\ref{k=pv-v-11})  which guarantee also the light-likeness of the momenta.
The  'energy variables' $\rho_i^{\#}$ are introduced to increase the (gauge) symmetry of the relation (\ref{k=pv-v-11}) to $SO(1,1)_i\times SO(8)_i$
(where the index $i$ is introduced to stress that, in the scattering problem,  each set of spinor frame variables is defined up to its 'own'  $SO(1,1)\times SO(8)$ gauge transformations). This makes possible to identify  $ v_{\alpha q(i)}^{\; -}$ with homogeneous coordinates of ${\bb S}^{8}$, (\ref{v-=S8}), which, in the light of the relation with light-like momenta (\ref{k=pv-v-11}), can be recognized as celestial sphere of a ten-dimensional observer
(of the $i$-th 10D observer) \cite{Galperin:1991gk,Delduc:1991ir}.

The name of spinor frame variables indicates that the above constraints (\ref{k=pv-v-11}) can be obtained from two statements:
i) that the variables  $v_{\alpha q}^{\; -}$ form a 16$\times$8 block of a $Spin (1,9)$ valued matrix
\footnote{Hence also the name of Lorentz harmonics \cite{Bandos:1990ji,Galperin:1991gk,Delduc:1991ir}.}
\begin{eqnarray}\label{harmV=D}
V_{\alpha}^{(\beta)}= \left( v_{\alpha \dot{q}}^{\; +} , v_{\alpha q}^{\; -}
  \right) \in Spin(1,9)\; , \qquad {q}=1,...,8\; ,  \qquad \dot{q}=1,...,8\; ,  \quad
\end{eqnarray}
which is called {\it spinor moving frame matrix}, and ii) that the light-like momentum $k_a$ of a massless 10D particle,   $k_ak^a=0$,
is related
to certain  vector from associated  $SO(1,9)$ valued matrix (moving  frame matrix)
\begin{eqnarray}\label{uaib=}
u_{ai}^{(b)}= \left(  \left(u_{a (i)}^{\#} +  u_{a i}^=\right)/2 , u_{a i}^I\, , \left(u_{a i}^\# - u_{a i}^=\right)/2\right) \quad \in \quad SO(1, D-1)\; \qquad
\end{eqnarray}
by (for further use, we restore the subscript $i=1,...,n$  here)
\begin{eqnarray}\label{kai=}
k_{ai} = \rho^{\#}_i u_{ai}^{=}\; . \qquad
\end{eqnarray}
The relation of moving frame (\ref{uaib=}) and spinor moving frame (\ref{harmV=D}) is  given by
\begin{eqnarray}\label{VGVt=G} V\sigma_b V^T =  u_b^{(a)} {\sigma}_{(a)}\, , \qquad V^T \tilde{\sigma}^{(a)}  V = \tilde{\sigma}^{b} u_b^{(a)}\;
 \, , \qquad \end{eqnarray}
 which can be easily recognized as conditions of Lorentz invariance of the generalized Pauli matrices written for a specific Lorentz rotation associated to the vector frame \footnote{This is the  Lorentz rotation from  a special coordinate system  in which
 $k_{(a)i} = \rho^{\#}_i (1,0,...,0,-1)$ to an arbitrary coordinate system  under consideration.}.

Eq. (\ref{uaib=})  implies the following properties of the frame vectors (or vector harmonics; these were called light-cone harmonic variables in \cite{Sokatchev:1985tc,Sokatchev:1987nk})
\begin{eqnarray}\label{uu=0}
&& u_a^{=}u^{a=}=0 \; , \qquad  \label{uu=2}
  u_a^{\#}u^{a\#}=0 \; , \qquad  u_a^{=}u^{a\#}=2 \; , \qquad \\ \label{uui=0} && u_a^{I}u^{a=}=0 \; , \qquad u_a^{I}u^{a\#}=0 \; , \qquad  u_a^{I}u^{aJ}=-\delta^{IJ} \; , \qquad
\end{eqnarray}

With an appropriate representation of sigma matrices, (\ref{VGVt=G}) implies
\begin{eqnarray}\label{u==v-v-}
 & u_a^= \sigma^a_{\alpha\beta}= 2v_{\alpha q}^{\;\; -} v_{\beta q}^{\;\; -}  \; , & \qquad
 u_a^= \delta_{{q}{p}} = v^-_{{q}} \tilde{\sigma}_{a}v^-_{{p}}  \qquad  \\
\label{v+v+=u++}
& v_{\dot{q}}^+ \tilde{\sigma}_{ {a}} v_{\dot{p}}^+ = \; u_{ {a}}^{\# } \delta_{\dot{q}\dot{p}}\; , & \qquad 2 v_{{\alpha}\dot{q}}^{\; \;+}v_{{\beta}\dot{q}}^{\;\; +}= {\sigma}^{ {a}}_{ {\alpha} {\beta}} u_{ {a}}^{\# }\; , \qquad \\
 \label{uIs=v+v-}
& v_{{q}}^- \tilde {\sigma}_{ {a}} v_{\dot{p}}^+=u_{ {a}}^{I} \gamma^I_{q\dot{p}}\; , &\qquad
 2 v_{( {\alpha}|{q} }^{\;\;\; -} \gamma^I_{q\dot{q}}v_{|{\beta})\dot{q}}^{\;\;\; +}= {\sigma}^{a}_{\alpha\beta} u_{ {a}}^{I}\; , \quad  \end{eqnarray}
where  $\gamma^I_{q\dot{p}}= \tilde{\gamma}{}^I_{\dot{p}q}$ with  $I=1,...,8$ are  $SO(8)$ Clebsch-Gordan coefficients,
 which obey
 $ \gamma^I\tilde{\gamma}{}^J+ \gamma^J\tilde{\gamma}{}^I= \delta^{IJ}I_{8\times 8}$ and $ \tilde{\gamma}{}^I\gamma^J+ \tilde{\gamma}{}^J\gamma^I= \delta^{IJ}I_{8\times 8}$.

The 10D spinor helicity variables $\lambda_{\alpha q \, i}= \sqrt{\rho^{\#}_i}v^{\; -}_{\alpha q \, i}$ were introduced in \cite{CaronHuot:2010rj} and used their to construct a Clifford superfield approach to superamplitude. The understanding of the Lorentz harmonic nature of spinor helicity variables from \cite{CaronHuot:2010rj}\footnote{This {\it helicity spinor--Lorentz harmonic} correspondence was also noticed in \cite{Uvarov:2015rxa} in a context of five dimensional field theories. } allowed us to construct the spinor helicity formalism for 11D supergravity \cite{Bandos:2016tsm}, simplify it for 10D SYM \cite{Bandos:2017eof} and propose two versions of  superamplitude formalism for  11D SUGRA and 10D SYM  \cite{Bandos:2016tsm,Bandos:2017eof,Bandos:2017zap}
(both simpler than the 10D Clifford superamplitude approach of \cite{CaronHuot:2010rj}).

Eqs. (\ref{uu=0}) and (\ref{uui=0}) follow from (\ref{u==v-v-})--(\ref{uIs=v+v-}). The relations (\ref{k=pv-v-11}) follow from (\ref{u==v-v-}) and (\ref{kai=}). What remains to comment is how the statement in (\ref{v-=S8}) occurs.

In distinction to $v_{\alpha p }^{\; -}$, the complementary harmonic variables $v_{\alpha \dot{q} }^{\; +} $ are not physical and serve as a set of reference spinors (a counterpart of 4D ${\mu}_{{\alpha}},\bar{\mu}_{\dot{\alpha}}$). This is reflected by $K_8$ gauge  symmetry  of the Lorentz harmonic description of  massless particles which acts on spinor frame as
\begin{eqnarray}\label{KD-2=v}
& K_{8}\; : \qquad v_{\alpha \dot{q} }^{\; +} \mapsto v_{\alpha \dot{q} }^{\; +} + \frac 1 2 K^{\# I}v_{\alpha p }^{\; -}\gamma^I_{p\dot{q}}\; , \qquad v_{\alpha {q} }^{\; -} \mapsto v_{\alpha {q} }^{\; -}  \; .   \qquad
\end{eqnarray}

The gauge symmetry  $\prod_i SO(1,1)_i\otimes SO(8)_i\subset\!\!\!\!\!\! \times K_{8i}$ make possible to identify the Lorentz  harmonics variables  $ (v^-_{\alpha q(i)}, v^+_{\alpha \dot{q}(i)})$  with generalized homogeneous coordinates of the coset isomorphic to the celestial sphere,
 \begin{eqnarray}\label{v-i=G-H}
 & \{ (v^-_{\alpha q(i)}, v^+_{\alpha \dot{q}(i)})\}    = \left(\frac {Spin(1,9)}{[SO(1,1)\otimes Spin(8)]\subset\!\!\!\!\times K_{8}}\right)_i={\bb S}^{8}_i \; . \qquad  \end{eqnarray}

 To make the statement in (\ref{v-i=G-H}) manifest, we can introduce an arbitrary reference spinor frame $(v^-_{\alpha q}, v^+_{\alpha \dot{q}})$ and then fix the  $\prod_i^\otimes SO(1,1)_i\otimes SO(8)_i\subset\!\!\!\!\!\! \times K_{8i}$ auxiliary gauge symmetries by representing the $i$-th spinor frame as \cite{Bandos:2017eof,Bandos:2017zap}:
\begin{eqnarray}\label{v-j=v-+Kv-}
& v_{\alpha qi}^{\; -} =
v_{\alpha  q}^{\; -}+ {1\over 2} K^{=I}_{i}  \gamma^I_{q\dot{q}} v_{\alpha  \dot{q}}^{\; +}
 \; ,  \qquad
v_{\alpha  \dot{q}i}^{\; +}=   v_{\alpha {\dot{q}}}^{\;+}
 \; .  \qquad
\end{eqnarray}
Then  eight variables $ K^{=I}_{i}  $ carrying the physical degrees of freedom in $v_{\alpha qi}^{\; -}$ can be identified
with (stereographic) projective coordinates of ${\bb S}^8$ sphere.

When calculating amplitudes with our spinor frame based spinor helicity formalism, it is often convenient and/or
 instructive to fix only the $\prod_i^\otimes K_{8i}$ symmetry and to write the variables of the i-th spinor frame as \cite{Bandos:2017eof,Bandos:2017zap}
\begin{eqnarray}\label{v-j=+Kv-}
& v_{\alpha qi}^{\; -} =e^{-\alpha_i} {\cal O}_{i{q}{p}}  \left(
v_{\alpha  p}^{\; -}+ {1\over 2} K^{=I}_{i}  \gamma^I_{p\dot{q}} v_{\alpha  \dot{q}}^{\; +}
\right) \; ,  \qquad
\label{v+j=+Kv-}
v_{\alpha  \dot{q}i}^{\; +}=  {\cal O}_{i\dot{q}\dot{p}}
e^{\; \alpha_{i}} v_{\alpha {\dot{p}}}^{\;+}
 \; ,  \qquad
\end{eqnarray}
where $ {\cal O}_{i{q}{p}}$ and $ {\cal O}_{i\dot{q}\dot{p}}$ are $SO(8)$ valued matrices 'parametrizing' the $SO(8)_i$ group, $ {\cal O}_{i{q}{p}}\gamma^I_{{q}\dot{q}}{\cal O}_{i\dot{q}\dot{p}} =  \gamma^J_{{q}\dot{q}}{\cal O}_{i}^{JI}$, and $\alpha_i$ are parameters of $SO(1,1)_i$.

{\bf 2.2} {\it The internal harmonics $(w^{A}_{qi}, \bar{w}{}_{Aqi})$ (\ref{wi=G-H=alt})}  are defined by their relation with a single set of reference internal frame variables $(w^{A}_{q}, \bar{w}{}_{Aq})$:
\begin{eqnarray}\label{bwi=bwOU}
\bar{w}_{{q}A\, i}= {\cal O}_{{q}{p}\, i} \bar{w}_{{p}B} \; e^{-i\beta _i} \, {\cal U}_{A\, i}^{\dagger\, B}\; , \qquad
   {w}_{q\, i}^{\; A} = {\cal O}_{{q}{p}\, i} {w}_{p}^{\; B} \;  e^{+i\beta _i}  {\cal U}_{B\, i}^{\;  A}\; , \qquad
  \\  \label{bwid=bwOU}
\bar{w}_{\dot{q}A\, i}= {\cal O}_{\dot{q}\dot{p}\, i} \bar{w}_{\dot{p}B} \; e^{i\beta _i} \, {\cal U}_{A\, i}^{\dagger\, B}\; , \qquad
   {w}_{\dot{q}\, i}^{\; A} = {\cal O}_{\dot{q}\dot{p}\, i} {w}_{\dot{p}}^{\; B} \;  e^{-i\beta_i}  {\cal U}_{B\, i}^{\;  A}\; , \qquad
   \\ \label{cUinOinU}
   {\cal U}_{A\, i}^{\dagger\, C} {\cal U}_{C\, i}^{\;\, B}=\delta_A{}^B\; \quad \Leftrightarrow \qquad
    {\cal U}_{B\, i}^{\;  A} \in  SU(4)\; . \qquad
\end{eqnarray}
They are needed to construct the complex fermionic coordinate $\eta^-_{Ai}= (\bar{\eta}{}^{-A}_i)^*$ with
$SU(4)_i$ index $A=1,...,4$, the arguments of the superamplitude  (\ref{cA=alt}), starting from a real fermionic coordinate $\theta^-_{qi}=(\theta^-_{qi})^*$ with
$SO(8)_i$ s-spinor index $q=1,...,8$,
\begin{eqnarray}\label{etaA=thbw}
\eta^-_{Ai}=  \theta^-_{qi} \bar{w}_{{q}A\, i}
\; . \qquad
\end{eqnarray}
To understand the origin of these two types of fermionic variables, one may turn to the quantization of massless superparticle. We refer to \cite{Bandos:2017eof} for details of that and only notice here that the 8-component real fermionic coordinate $\theta^-_{qi}$ in its turn is composed of the real MW spinor fermionic coordinate
$\theta^\alpha$ and the corresponding spinor frame variable $v_{\alpha q i}^{\; -}$,
\begin{eqnarray}\label{th-q=thv}
\theta^-_{qi} = \theta^\alpha_i v_{\alpha q i}^{\; -}
\; . \qquad
\end{eqnarray}
Eqs. (\ref{etaA=thbw}) and (\ref{th-q=thv}) implies that the complex fermionic coordinate is actually constructed from the real MW spinor and complex harmonics:
$
\eta^-_{Ai}=  \theta^\alpha_i v_{\alpha A i}^{\; -}$, $\; v_{\alpha A i}^{\; -}:=  v_{\alpha q i}^{\; -} \bar{w}_{{q}A\, i}$. This is a manifestation of a more general fact that the amplitude (\ref{cA=alt}) can be considered as a function \begin{eqnarray}\label{cA=alt2}
{\cal A}_n={\cal A}_n(\{ \rho^{\#}_i, v_{\alpha A i}^{-}, v_{ Ai}^{-\alpha}; \eta^-_{Ai}\})\; \qquad
\end{eqnarray}
of complex spinor frame variables
$ v_{\alpha A}^{-}:= v_{\alpha q}^{-} \bar{w}_{qA}$ and $v_{ A}^{-\alpha}:= v_{\dot{q}}^{-\alpha} \bar{w}_{\dot{q}A} $ parametrizing the coset $\frac{Spin(1,9)}{[SU(4)\otimes U(1)\otimes SO(1,1)]\subset\!\!\!\! \times K_{8}}$
(see \cite{Bandos:2017zap} for further details).

The analytic superamplitudes (\ref{cA=alt}) ((\ref{cA=alt2})) do not carry indices,
are Lorentz invariant,  invariant under
$\prod\limits_{i=1}^n [ SO(1,1)_i\otimes SO(8)_i\otimes SU(4)_i]$ ($\prod\limits_{i=1}^n [ SO(1,1)_i\otimes SU(4)_i]$)
and covariant under $\prod\limits_i SO(2)_i=U(1)_i$ symmetry transformations.

\if{}
As described in \cite{Bandos:2017eof,Bandos:2017zap}, there are two ways of gauge fixed covariant quantizations, one of which leads to the one-particle counterpart of the analytic superamplitude, which we are discussing now, and the other to the constrained on-shell superfields
\cite{Bandos:2017eof}, the counterpart of the constrained superamplitudes which will be briefly discussed below.
\fi

 {\bf 2.3.} Eqs. (\ref{etaA=thbw}) and  (\ref{th-q=thv}) reveal {\it a non-manifest difference of the fermionic arguments  of $D=4$ and $D=10$ superamplitudes}. The former, $\eta_{Ai}$, carry the index of the same $SU(4)$ group, the R-symmetry group of ${\cal N}=4$ 4D supersymmetry, while the latter, $\eta^-_{Ai}$, carry the indices of different $SU(4)_i$ gauge symmetry groups, the transformations of which are used as identification relations allowing us to treat the internal harmonics as coordinate of the coset (\ref{wi=G-H=alt}).
The matrices ${\cal U}_{B\, i}^{\;  A} \in  SU(4)$ in  (\ref{bwi=bwOU})--(\ref{cUinOinU}) are 'bridges' between these  $SU(4)_i$'s and the $SU(4)$ gauge symmetry used to define the reference internal frame
 matrix
$ \{ w^{A}_{q}, \bar{w}{}_{Aq}\}$ and its counterpart with c-spinor indices
$\{ w^{A}_{\dot{q}}, \bar{w}{}_{A\dot{q}}\}$. The 8$\times$4 complex conjugate blocks of these matrices obey
$w_{q}{}^A \bar{w}_{pA}+ \bar{w}_{qA}w_p{}^A =\delta_{qp}$ and
\begin{eqnarray} \label{bww=1}
&& \bar{w}_{qB}w_{q}{}^A =\delta_B{}^A\; , \qquad w_{q}{}^A w_{q}{}^B =0 \; , \qquad \bar{w}_{qA} \bar{w}_{qB} =0\; . \qquad
\end{eqnarray}
Due to specific properties of $SO(8)$, these constraints actually imply that $\{ w^{A}_{q}, \bar{w}{}_{Aq}\}$  and  $\{ w^{A}_{\dot{q}}, \bar{w}{}_{A\dot{q}}\}$ are formed from the linear combinations of the columns of the $Spin(8)$ valued matrices $w_{q}^{(p)}$ and $w_{\dot{q}}^{(\dot{p})} $ related to 8v- representation of the SO(8) reference internal frame
\begin{eqnarray}\label{UinSO8}
& U_I^{(J)}= \left(U_I{}^{\check{J}}, U_I{}^{(7)}, U_I{}^{(8)}\right)= \left(U_I{}^{\check{J}}, \frac 1 2 \left( U_I+ \bar{U}_I\right), \frac 1 {2i} \left( U_I- \bar{U}_I \right)\right)  \in \, SO(8)
 \quad
\end{eqnarray}
by $\gamma^I_{q \dot{p}} U_I^{(J)}= w_q^{(p)}\gamma^{(J)}_{(p) (\dot{q})}  w_{\dot{p}}^{(\dot{q})}$. This includes
the following factorization relations for two null-vectors of the internal 8v-frame, $U_I$ and $\bar{U}_I=({U}_I)^*$,
\begin{eqnarray} \label{Ug8=bww}
U\!\!\!\!/{}_{q \dot{p}}:=  \gamma^I_{q \dot{p}} U_I = 2  \bar{w}_{{q}A} w_{\dot{p}}^A  \; , \qquad
 \bar{U}\!\!\!\!/{}_{q \dot{p}}:= \gamma^I_{q \dot{p}} \bar{U}_I = 2   w_q^{A} \bar{w}_{\dot{p}A} \; . \qquad
\end{eqnarray}
$U_IU_I=0$ and $ \bar{U}_I  \bar{U}_I =0$ follow from Eqs.  (\ref{Ug8=bww}) and (\ref{bww=1}).

{\bf 3.} {\it Analytic superamplitude from constrained superamplitude}. Just the above complex null vectors are used to construct the analytic superamplitude (\ref{cA=alt}) from the basic constrained superamplitude  of 10D SYM theory
$ {\cal A}_{I_1... I_n} (\{ \rho^\#_{i}, v_{\alpha qi}^{\; -}; \theta_{ qi}^{-}\})$ \cite{Bandos:2017eof}. This carries $n$ 8v- indices of $SO(8)_i$ 'small groups' of the light-like momenta (\ref{kai=}), depends on real spinor frame variables $v_{\alpha qi}^{\; -}$, densities $\rho^\#_{i}$ and real 8s-spinor  fermionic variables $\theta_{ qi}^{-}$, and obeys the set of superfield equations
\begin{eqnarray}\label{D+qAI=GA}
D^{+(j)}_{{q}_j} {\cal A}^{(n)}_{I_1... I_j...I_n}
=   2 \rho^{\#}_{j}\, \gamma^{I_j}_{q_j\dot{q}_j} {\cal A}^{(n)}_{I_1... I_{j-1}\dot{q}_j I_{j+1}... I_n}\; .
\end{eqnarray}
Here
$ D^{+(j)}_{{q}} = \frac {\partial} {\partial\theta^-_{{q}j}} + 2 \rho^{\#}_{j}\theta^-_{{q}j}$
and the fermionic constrained superamplitude ${\cal A}^{(n)}_{I_1... I_{j-1}\dot{q}_j I_{j+1}... I_n}$ is defined by gamma trace part of the same equation (\ref{D+qAI=GA}) (see \cite{Bandos:2017eof} for further details).

The analytic superamplitude (\ref{cA=alt}) is expressed in terms of constrained superamplitude $ {\cal A}_{I_1... I_n} (\{ \rho^\#_{i}, v_{\alpha qi}^{\; -}; \theta_{ qi}^{-}\})$ obeying (\ref{D+qAI=GA}) by \cite{Bandos:2017zap} ($U_{Ii}:= U_J {\cal O}_{JI\, i}e^{-2i\beta_i}$)
\begin{eqnarray} \label{cA=etcAU}
&& {\cal A}_n (\{ \rho^\#_{i}, v_{\alpha qi}^{\; -}; w_i, \bar{w}_i; \eta_{A i}\}) = \nonumber \\  && =e^{^{-2\sum_j  \rho^\#_{j} \eta^-_{Bj}  \bar{\eta}^{-B}_j }}
 U_{I_1 1}\ldots U_{I_n n}\; {\cal A}_{I_1... I_n} (\{ \rho^\#_{i}, v_{\alpha qi}^{\; -}; \; \eta^-_{Ai}  {w}^{\; A}_{qi} +   \bar{\eta}^{-A}_i \bar{w}_{q{A}i}\}) . \qquad
\end{eqnarray}

{\bf 4.} We would like to conclude this contribution by presenting the gauge fixed version of the basic {\it 3--point superamplitude of 10D SYM}, the counterpart of the 4D $\overline{MHV}$ amplitude (\ref{cA3g=bMHV}). In the gauge
(\ref{v-j=+Kv-}), (\ref{bwi=bwOU}) this reads
\begin{eqnarray}\label{cA3g=10DK}
 {\cal A}_3^{D=10\; SYM} &=&\frac 1 2  {{\cal K}{}^{==I}U_I} \; e^{-2i(\beta_1+\beta_2+\beta_3)}\, \delta^4 \left( \tilde{\rho}{}^{\#}_{1} \tilde{\eta}^-_{1A}   + \tilde{\rho}{}^{\#}_2 \tilde{\eta}^-_{2A}  + \tilde{\rho}{}^{\#}_3\tilde{\eta}^-_{3A}  \right)\; ,
 \qquad
  \end{eqnarray}
where $U_I$ is the null-vector from the reference internal frame (\ref{UinSO8}), (\ref{Ug8=bww}),
\begin{eqnarray}
 \label{tr=} \tilde{\rho}{}^{\#}_{i}:= {\rho}{}^{\#}_{i} e^{-2\alpha_i}\, \qquad \tilde{\eta}{}^-_{A i}:={\eta}{}^-_{B i}
 e^{\alpha_i+i\beta_i} {\cal U}_{A\, i}^{\; B}\; ,
 \qquad
  \end{eqnarray}   with $\alpha_i$, $\beta_i$ and ${\cal U}_{A\, i}^{\; B}$ defined in (\ref{v-j=+Kv-}) and (\ref{bwi=bwOU}),  and
the complex null-vector ${\cal K}{}^{==I}$ is defined by \begin{eqnarray}
 \label{Kij=q}
 &  \frac {{K}{}^{=I}_{32}}{\tilde{\rho}^\#_1}= \frac {{K}{}^{=I}_{21}}{\tilde{\rho}^\#_3}= \frac {{K}{}^{=I}_{13}}{\tilde{\rho}^\#_2} =: {\cal K}^{==I}\; , \qquad
 {K}{}^{=I}_{ij}={K}{}^{=I}_{i}-{K}{}^{=I}_{j}\; .
\end{eqnarray}
First and second  equalities in (\ref{Kij=q}) follow from the momentum conservation conditions in 3-particle process. These  requires ${\cal K}^{==I}$ to be complex and nilpotent, $ {\cal K}^{==I}{\cal K}^{==I}=0$. The use of the on-shell amplitudes dependent on deformed complex light-like momenta to calculate higher $n$ amplitudes of particles with real, physical light-like momenta  is a characteristic property of the BCFW approach \cite{Britto:2005fq,ArkaniHamed:2008gz}. The candidate 
BCFW recurrent relations for constrained superamplitudes of 10D SYM and 11D SUGRA can be found in \cite{Bandos:2017eof} and \cite{Bandos:2016tsm}. The structure of the BCFW deformation of complex spinor frame variables relevant for the calculation of the analytic 10D and 11D superamplitudes was discussed in \cite{Bandos:2017zap}.

{\bf Acknowledgments}.  This work has been supported in part by the
Spanish Ministry of Economy, Industry and Competitiveness (MINECO) grants FPA 2015-66793-P, which is partially financed with FEDER/ERDF fund of EU,
by the Basque Government Grant IT-979-16, and the Basque Country University program UFI 11/55.

The author is grateful to Theoretical Department of CERN (Geneva, Switzerland),
to the Galileo Galilei Institute for Theoretical Physics and the INFN (Florence, Italy), as well as to the
the organizers of the GGI workshop ''Supergravity: what next?'', and especially to  Antoine Van Proeyen,
for the hospitality and partial support of his visits at certain stages of this work.


Many thanks to the organizers of SQS 2017 conference, and especially to Zhenya Ivanov and Sergei Fedoruk, for their kind hospitality in Dubna.

\end{document}